\documentclass[aps,prl,twocolumn,floatfix,showpacs,preprintnumbers,amsmath,amssymb,10pt]{revtex4-1}
\usepackage[dvips]{graphics}

\usepackage{psfrag}
\usepackage{graphicx}
\usepackage{trfsigns}
\usepackage{amssymb}
\usepackage[usenames,dvipsnames]{color}
\usepackage{dashrule}
\usepackage{textgreek}
\usepackage[english]{babel}
\usepackage{contour}
\usepackage{upgreek,bm}
\usepackage{color}
\usepackage{dsfont}
\usepackage{soul}
\usepackage{braket}
\usepackage{cancel}
\usepackage{subfigure}
\usepackage{overpic}

\definecolor{dred}{rgb}{.6,.0,0.}
\definecolor{dblue}{rgb}{.0,.0,0.6}

\renewcommand{\vec}[1]{\mathbf{#1}}

\newcommand{\tens}[1]{\mbox{\textsf{\textbf{#1}}}}

\newcommand{\sprod}{\!\cdot\!}

\newcommand{\dif}{\mathrm{d}}
\newcommand{\mi}{\textrm{i}} 
\newcommand{\me}{\mathrm{e}}

\begin{document}

\title{Purcell--Dicke effect for an atomic ensemble near a surface}

\author{Sebastian Fuchs$^1$}
\author{Stefan Yoshi Buhmann$^{1,2}$}
\affiliation{$^1$ Physikalisches Institut, Albert-Ludwigs-Universit\"at Freiburg, Hermann-Herder-Stra{\ss}e 3, 79104 Freiburg, Germany\\
$^3$ Freiburg Institute for Advanced Studies, Albert-Ludwigs-Universit\"at Freiburg, Albertstra{\ss}e 19, 79104 Freiburg, Germany}

\date{\today}

\begin{abstract}
The emission by an initially completely inverted atomic ensemble in the long-wavelength regime is simultaneously enhanced by both collective effects (Dicke effect) and dielectric environments (Purcell effect), thus giving rise to a combined Purcell--Dicke effect. We study this effect by treating the ensemble of $N$ atoms as a single effective $N+1$-level `Dicke atom' which couples to the environment-assisted quantum electrodynamic field. We find that an environment can indeed alter the superradiant emission dynamics, as exemplified using a perfectly conducting plate. As the emission acquires an additional anisotropy in the presence of the plate, we find an associated resonant Casimir--Polder potential for the atom that is collectively enhanced and that exhibits a superradiant burst in its dynamics. An additional tuneability of the effect is introduced by applying an external driving laser field.
\end{abstract}

\maketitle
Recent studies of atom-light interactions in non-trivial environments have increasingly focused on collective effects \cite{Mlynek:2014, Stehle:2014}. In such efforts, two very distinct phenomena can potentially lead to a complex interplay: the Casimir--Polder potential of a single atom near a surface mediated by the vacuum fluctuations \cite{Casimir_Polder:1948} and the associated change of the atom's spontaneous emission rate due to the surface's presence (Purcell effect \cite{Purcell:1946}), and the collective enhancement of radiation in an atomic ensemble (Dicke effect \cite{Dicke:1954}).

The Casimir--Polder potential and its associated force are part of the field of dispersion forces, which are pure quantum forces in a sense that they stem from zero-point fluctuations of the quantized electromagnetic field. There are a variety of theoretical frameworks to describe the Casimir--Polder potential \cite{Barton:1970, Wylie:1984}. We use macroscopic quantum electrodynamics (QED), cf. e.g. Ref.~\cite{Scheel:2008}, an extension of vacuum QED incorporating material properties by macroscopic response functions such as permittivity or permeability. Experimentally, Casimir--Polder forces can be made visible directly via the change of the atomic motion under the influence of the potential, e.g. by measuring the deflection angle of atoms passing a macroscopic object such as a V-shaped cavity \cite{Sukenik:1993}, or in an indirect approach using spectroscopy to detect the shift of the atomic transition frequency \cite{Failache:1999}.

Originally defined as the enhancement of the rate of spontaneous decay of an atom coupled to a single-mode resonant cavity \cite{Purcell:1946, Fox_Book}, the term Purcell effect can also be extended to a situation where the atom is in vicinity to a single surface providing a large density of states and thus an enhancement of the radiative decay \cite{Iwase:2010, Cohen-Tannoudji_Book2}. This has been exploited, e.g. by increasing light emission of quantum wells in InGaN light-emitting diodes by means of surface plasmons \cite{Okamoto:2004}.

Superradiance is another effect which enhances light emission, in this case by a collective mechanism in an ensemble of atoms \cite{Haroche:1982}. The characteristics of the emerging superradiant peak for a total number of atoms $N$ are the proportionality of its height to $N^2$ and its width to $1/N$. The first experimental verification of superradiance was realized in an optically pumped hydrogen fluoride gas \cite{Skribanowitz:1973}. After this first detection, superradiance was studied for several systems, e.g. in quantum dots \cite{Scheibner:2007}, single diamond nanocrystals \cite{Bradac:2017}, Rydberg atoms \cite{Grimes:2017}, or artificial atoms in a cavity \cite{Mlynek:2014}. As a further development, collective effects have recently been demonstrated to impact atom--light interactions near surfaces \cite{Stehle:2014}. There, by studying the cooperative coupling of ultracold atoms to surface plasmons propagating on a plane gold surface a Purcell enhancement of the atomic fluorescence caused by the surface plasmons was found.

\begin{figure}[t!]
	\centering
		\includegraphics[width=\columnwidth]{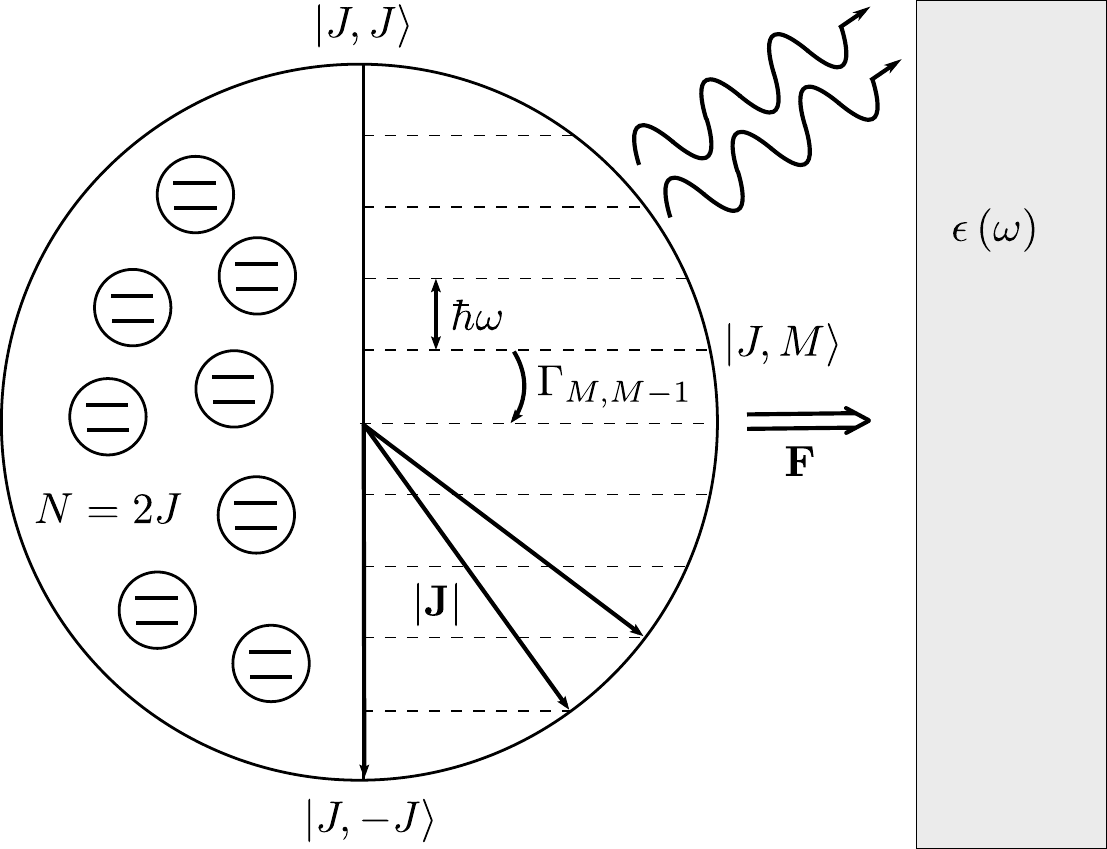}
	\caption{Purcell--Dicke effect: $N$ identical two-level atoms form a `Dicke atom' with equidistant Dicke states $\ket{J,M}$ \eqref{eq:Dicke State} separated from each other by $\hbar \omega$, and decay rates to the neighboring states $\Gamma_{M,M-1}$. The presence of the surface with permittivity $\epsilon \left( \omega \right)$ alters the radiation properties of the `Dicke atom' causing an attractive Casimir--Polder force.}
	\label{fig:Large Spin}
\end{figure}

This Letter reports on the superradiant intensity of atomic emitters in arbitrary structured environments, illustrated for the example of a perfectly conducting planar surface (see Fig.~\ref{fig:Large Spin}). In addition, we demonstrate how the Dicke effect is manifested in an enhanced Casimir--Polder interaction between the atomic ensemble and the surface. Note that Ref.~\cite{Sinha:2018} presents calculations of the collective Casimir--Polder force in the same spirit based on a complementary approach involving master equations.

We are going to study the influence of a nearby surface on the superradiant emission burst of an atomic cloud and derive the associated collective Casimir--Polder potential due to photon recoil. The Hamiltonian of this system $\hat{H} = \hat{H}_{\textrm{F}} + \hat{H}_{\textrm{A}} + \hat{H}_{\textrm{AF}}$ consists of the surface-assisted field contribution $\hat{H}_{\textrm{F}}$, the collective atomic Hamiltonian $\hat{H}_{\textrm{A}}$ and the interaction term coupling the atoms to the electromagnetic field $\hat{H}_{\textrm{AF}}$. In the framework of macroscopic quantum electrodynamics (QED) \cite{Buhmann_Book_1, Buhmann_Book_2}, spontaneously fluctuating noise currents are described by polariton-like annihilation and creation operators $\hat{\vec{f}} \left( \vec{r}, \omega \right)$ and $\hat{\vec{f}}^{\dagger} \left( \vec{r}, \omega \right)$, which form the Hamiltonian of the medium-assisted electromagnetic field $\hat{H}_{\textrm{F}}$. The frequency components of the electric field $\hat{\vec{E}} \left( \vec{r}, \omega \right)$ are given by \cite{Buhmann:2004} $\hat{\vec{E}} \left( \vec{r}, \omega \right) = \int \dif^3 r' \tens{G} \left( \vec{r}, \vec{r}', \omega \right) \sprod \hat{\vec{f}} \left( \vec{r}', \omega \right)$, where the classical Green's tensor $\tens{G} \left( \vec{r}, \vec{r}', \omega \right)$ is the formal solution of the Helmholtz equation for the electromagnetic field \cite{Buhmann:2004}.

The electric field $\hat{\vec{E}} \left( \vec{r}_i \right)$ at the position $\vec{r}_i$ of each atom $i$ couples to the respective dipole moments $\hat{\vec{d}}_i$ according to the multipolar coupling, yielding the interaction Hamiltonian $\hat{H}_{\textrm{AF}}$
%%%
\begin{equation}
\hat{H}_{\textrm{AF}} = - \sum\limits^N_{i=1} \hat{\vec{d}}_i \sprod \hat{\vec{E}} \left( \vec{r}_i \right) = - \left( \sum\limits^N_{i=1} \hat{\vec{d}}_i \right) \sprod \hat{\vec{E}} \left( \vec{r}_{\textrm{A}} \right),
\label{eq:Interaction Hamiltonian}
\end{equation}
%%%
where $N$ is the total number of atoms in the cloud.
According to the Dicke model \cite{Dicke:1954}, the atoms are assumed to be motionless and confined to a volume much smaller than the wavelength $\lambda$ of the applied or emitted fields. In essence all atoms then feel the same average electromagnetic field $\hat{\vec{E}} \left( \vec{r}_{\textrm{A}} \right)$.

As outlined in Ref.~\cite{Haroche:1982}, one can then establish the Dicke states as symmetric eigenstates of the atomic ensemble. All atoms are considered as identical two-level systems with an excited state $\ket{e}$ and a ground state $\ket{g}$ separated by an energy $\hbar \omega_{\textrm{A}}$ and coupled by the single-atom dipole moment $\vec{d} = \bra{e} \hat{\vec{d}}_i \ket{g}$. Initially, the atoms are prepared in the maximally excited state
%%%
\begin{equation}
\ket{\psi \left( 0 \right)} = \ket{e,e,...,e}.
\label{eq:Initial State}
\end{equation}
%%%
Here, we assume that the atoms interact only with the electromagnetic field and atomic collisions or other relaxation processes are discarded. All atomic states being involved in the subsequent evolution have to be invariant with respect to an exchange of any two atoms. This property is represented by a symmetric superposition of $N$ spin-$1/2$ states, which is an eigenstate of the angular momentum operator $\hat{J}$ at its maximal eigenvalue $J = N/2$. These $N+1$ collective Dicke states can be obtained by successive application of the symmetric collective deexcitation operator to the initial state \eqref{eq:Initial State}
%%%
\begin{equation}
\ket{J,M} = \sqrt{\frac{\left( J+M \right)!}{N! \left( J-M \right)!}} \left( \sum\limits^N_{i=1} \hat{\sigma}^-_i \right)^{\left( J-M \right)} \ket{e,e,...,e}
\label{eq:Dicke State}
\end{equation}
%%%
with $-J \leq M \leq J$. A general Dicke state $\ket{J,M}$ for $N$ atoms can be represented as
%%%
\begin{equation}
\ket{J,M} = \hat{\mathcal{S}} | \underbrace{e,e,...,e}_{J+M},\underbrace{g,g,...,g}_{J-M} \rangle,
\end{equation}
%%%
where there are $\left( J+M \right)!$ possibilities to arrange the excited atoms and $\left( J-M \right)!$ possibilities for the ground-state atoms. Making use of the normalized symmetrization operator $\hat{\mathcal{S}}$ the totally symmetrical state $\ket{J,M}$ \eqref{eq:Dicke State} has
%%%
\begin{equation}
\binom{N}{J+M} = \frac{N!}{\left( J-M \right)! \left( J+M \right)!}
\label{eq:Number of Permutations Dicke}
\end{equation}
%%%
distinct contributions. The square root of this expression serves as normalizing factor for the completely symmetric state $\ket{J,M}$.

Collective operators can be introduced by
%%%
\begin{equation}
\hat{J}^{\pm} = \sum\limits^N_{i=1}{\hat{\sigma}^{\pm}_i}; \; \; \hat{J}^z = \sum\limits^N_{i=1}{\hat{\sigma}^z_i}
\label{eq:Collective Operators}
\end{equation}
%%%
and are analogous to the operators of angular momentum with $J=N/2$. Using Eq.~\eqref{eq:Collective Operators}, the atomic Hamiltonian $\hat{H}_{\textrm{A}}$ can be written as $\hat{H}_{\textrm{A}} = \frac{1}{2} \hbar \sum_{i=1} \omega_{\textrm{A}} \hat{\sigma}^z_i = \frac{1}{2} \hbar \omega_{\textrm{A}} \hat{J}^z$. Making use of the collective operators \eqref{eq:Collective Operators} the atomic cloud of identical two-level atoms may be regarded as one single `Dicke atom' having dipole transitions only between neighboring states, which are all separated by $\hbar \omega_{\textrm{A}}$. Figure \ref{fig:Large Spin} depicts the Dicke states $\ket{J,M}$ \eqref{eq:Dicke State} as eigenstates of angular momentum. The interaction Hamiltonian $\hat{H}_{\textrm{AF}}$ can be expressed as $\hat{H}_{\textrm{AF}} = - ( \hat{J}^+ + \hat{J}^- ) \vec{d} \sprod \hat{\vec{E}} \left( \vec{r}_{\textrm{A}} \right)$ with the single-atom dipole moment $\vec{d}$.

One can then define the rate of photon emission in state $\ket{J,M}$ decaying to state $\ket{J,M-1}$ in the Dicke picture based on the decay rate of a single atom \cite{Agarwal:1975, Buhmann_Book_2} as
%%%
\begin{equation}
\Gamma_{M,M-1} =  \frac{2 \mu_0}{\hbar} \omega^2_{\textrm{A}} \vec{d}_{M,M-1} \sprod \textrm{Im} \tens{G} \left( \vec{r}_{\textrm{A}}, \vec{r}_{\textrm{A}}, \omega_{\textrm{A}} \right) \sprod \vec{d}_{M-1,M}.
\label{eq:Dicke Decay Rate}
\end{equation}
%%%
Next, we express the collective dipole moments $\vec{d}_{M,M-1}$ and $\vec{d}_{M-1,M}$ in Eq.~\eqref{eq:Dicke Decay Rate} by single-atom dipole moments $\vec{d}$, which contribute to the single-atom decay rate $\Gamma$. The atom-field coupling \eqref{eq:Interaction Hamiltonian} for $N$ dipole operators contains $N$ single-atom dipole operators, which need to act between a ground state $\ket{g}$ and an excited state $\ket{e}$ at the same position to have a dipole moment with non-zero contribution, whereas the $N-1$ atoms at the other positions have to be in the same state. This results in a number of permutations given by the product of $N$ and the number of permutations for $J+M-1$ excited atoms in $N-1$ atoms in total, cf. Eq.~\eqref{eq:Number of Permutations Dicke}, yielding a general expression for the Dicke state dipole moment
%%%
\begin{equation}
\vec{d}_{M,M-1}= \sqrt{\left( J+M \right) \left( J-M+1 \right)} \vec{d}.
\label{eq:Scaling Properties Dipole Moment}
\end{equation}
%%%
The decay rate of the Dicke state $\ket{J,M}$ \eqref{eq:Dicke Decay Rate} contains the two dipole moments $\vec{d}_{M,M-1}$ \eqref{eq:Scaling Properties Dipole Moment} and $\vec{d}_{M-1,M}$, resulting in an expression for the decay rate of $\Gamma_{M,M-1} = \left( J+M \right) \left( J-M+1 \right) \Gamma$, which is formally identical to the free-space result from Ref.~\cite{Haroche:1982}. These scaling properties represent an intrinsic connection between Dicke states \eqref{eq:Dicke State} and decay rates \eqref{eq:Dicke Decay Rate} and is the fundamental difference between one single atom with $N$ quantum states and a collection of $N$ atoms in the Dicke approximation. As we will see, this further leads to a superradiance-like scaling behavior of the collective Casimir--Polder potential for the atomic cloud.

As mentioned above, the surface's presence modifies the collective decay rate \eqref{eq:Dicke Decay Rate} by a Purcell factor $F_{\textrm{P}}$
%%%
\begin{equation}
\Gamma_{M,M-1} = \left( J+M \right) \left( J-M+1 \right) F_{\textrm{P}} \Gamma^{(0)}.
\end{equation}
%%%
This can be seen by decomposing the Green's tensor into $\tens{G} \left( \vec{r}_{\textrm{A}}, \vec{r}_{\textrm{A}}, \omega_{\textrm{A}} \right) = \tens{G}^{(0)} \left( \vec{r}_{\textrm{A}}, \vec{r}_{\textrm{A}}, \omega_{\textrm{A}} \right) + \tens{G}^{(1)} \left( \vec{r}_{\textrm{A}}, \vec{r}_{\textrm{A}}, \omega_{\textrm{A}} \right)$ with bulk part $\tens{G}^{(0)} \left( \vec{r}_{\textrm{A}}, \vec{r}_{\textrm{A}}, \omega_{\textrm{A}} \right)$ responsible for the free-space decay rate
%%%
\begin{equation}
\Gamma^{(0)} = \frac{\omega^3_{\textrm{A}} \left| \vec{d} \right|^2}{3 \pi \epsilon_0 \hbar c^3}
\label{eq:Free Space Decay Rate}
\end{equation}
%%%
and scattering part $\tens{G}^{(1)} \left( \vec{r}_{\textrm{A}}, \vec{r}_{\textrm{A}}, \omega_{\textrm{A}} \right)$ yielding the Purcell factor
%%%
\begin{equation}
F_{\textrm{P}} = 1 + \frac{6 \pi c}{\omega \left| \vec{d} \right|^2} \vec{d} \sprod \textrm{Im} \tens{G}^{(1)} \left( \vec{r}_{\textrm{A}}, \vec{r}_{\textrm{A}}, \omega_{\textrm{A}} \right) \sprod \vec{d}^*.
\end{equation}
%%%
At this point, it is worth checking the long-wavelength assumption of the Dicke model. To this end, we consider two atoms to be located at slightly different positions. Beside the symmetric superradiant state \eqref{eq:Dicke State}, an anti-symmetric subradiant state, which would be the singlet state if there were only two atoms, will emit radiation in this case. As shown in Ref.~\cite{Dzsotjan:2010} the decay process is governed by a joint decay rate of the two atoms $i=1$ and $i=2$: $\Gamma^{12}=\frac{2 \mu_0 \omega^2_{\textrm{A}}}{\hbar} \vec{d} \sprod \textrm{Im} \tens{G} \left( \vec{r}_1, \vec{r}_2, \omega_{\textrm{A}} \right) \sprod \vec{d}^*$.

To see if the conditions of superradiance hold, we need to check if the decay rate $\Gamma_{M=1,M=0} = \Gamma + \Gamma^{12}$ of the superradiant two-atom state is indeed equal to $2 \Gamma$ as suggested by Eq.~\eqref{eq:Dicke Decay Rate}. This is the case if $F=\Gamma^{12}/\Gamma \simeq 1$. We display this superradiance fidelity for two atoms near a perfectly conducting surface \cite{Palacino:2017} where we fix the position of one atom and vary the position of the other. One observes that the fidelity $F$ is indeed close to unity in a corridor around atom 1. This anisotropy of the superradiance region is induced by the presence of the surface, as seen by comparison with the free-space case.

\begin{figure}[t!]
	\centering
		\includegraphics[width=\columnwidth]{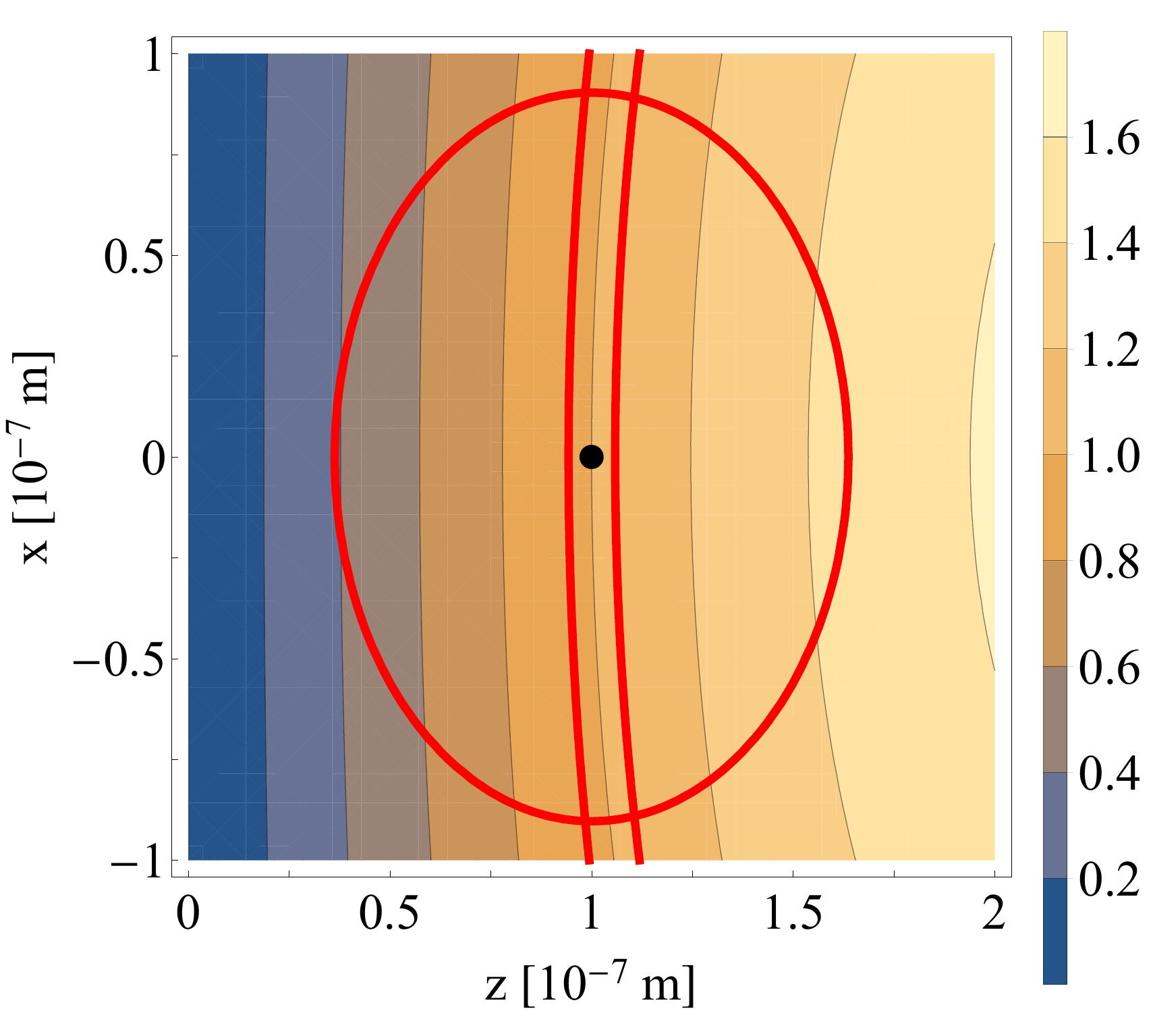}
	\caption{Superradiance fidelity $F=\Gamma^{12}/\Gamma$ for two atoms near a perfectly conducting plate. One atom is located at $x_{\textrm{A}} = 0, \; z_{\textrm{A}} = 10^{-7} \: \textrm{m}$. The dipole moment is polarized along the $x$-direction. $\Gamma^{12}$ is composed of the bulk rate $\Gamma^{12 (0)}$ and the scattered rate from an image dipole $\Gamma^{12 (1)}$ and thus vanishes at the boundary $z=0$. We indicate regions where $F= \left( 100 \pm 5 \right) \%$ (red corridor). For comparison, we also display the fidelity in free space $F^{(0)}= \Gamma^{(0)}_{12}/\Gamma^{(0)} = \left( 100 \pm 5 \right) \%$ (red circle). The shape and width of the corridor depend on the orientation of the dipole.}
	\label{fig:Decay_Rate}
\end{figure}

\begin{figure}[t!]
	\centering
		\includegraphics[width=\columnwidth]{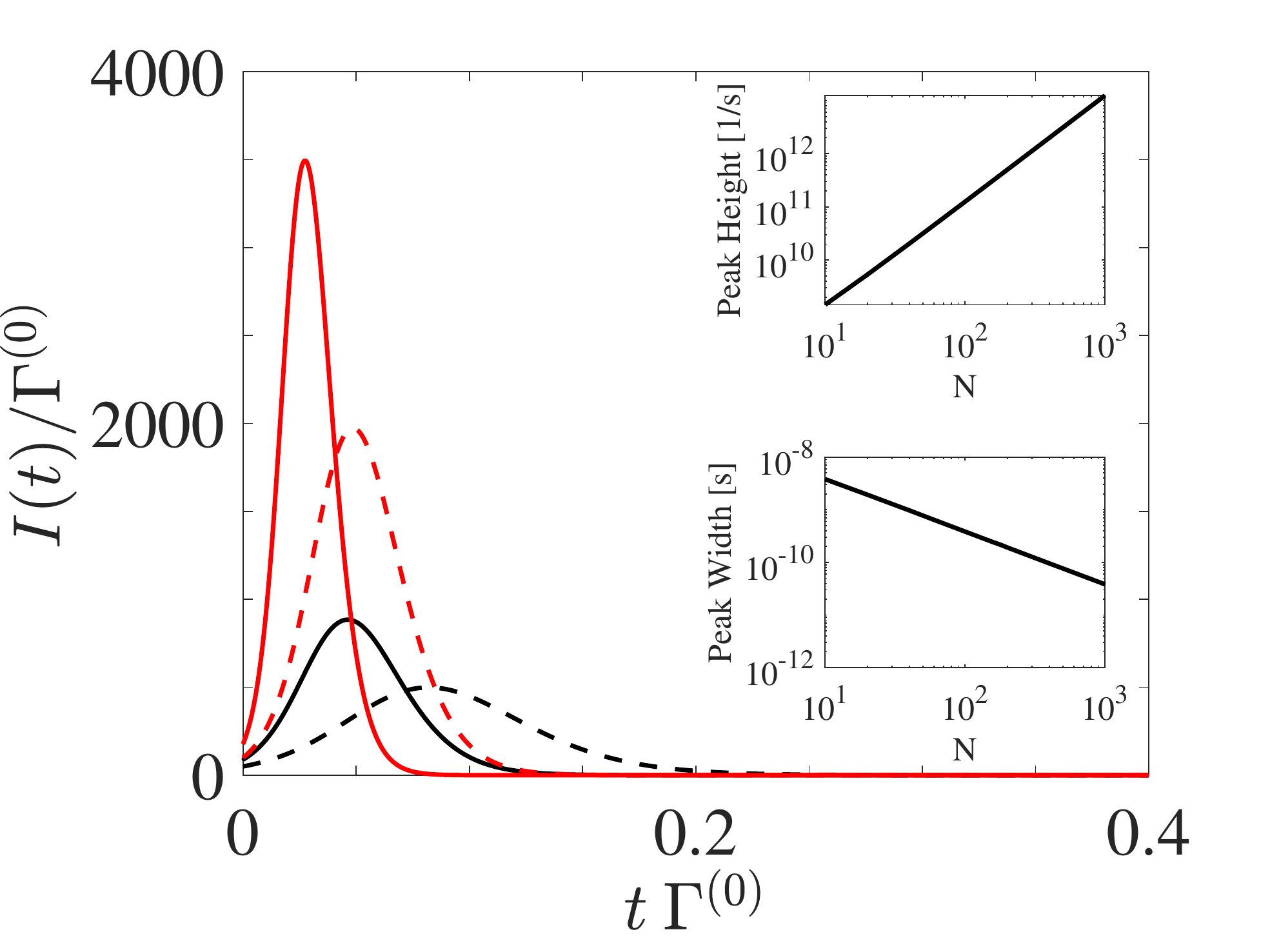}
	\caption{Total photon emission $I(t)$ \eqref{eq:Intensity} scaled by the free-space decay rate $\Gamma^{(0)}$ \eqref{eq:Free Space Decay Rate} for $N=50$ (black curves) and $N=100$ atoms (red curves) under the influence of a perfectly conducting surface. All atoms are located at $x_{\textrm{A}} = 0, \; z_{\textrm{A}} = 10^{-7} \: \textrm{m}$ and are polarized along the $z$-direction. The dashed lines show the respective photon emission for $N=50$ and $N=100$ atoms in the absence of the surface. The insets show the peak height and the full width at half maximum (FWHM) as a function of the number of atoms $N$. We obtain the relations: peak height $\propto N^{1.979}$ and FWHM $\propto N^{-1.001}$.}
	\label{fig:Intensity}
\end{figure}

In the following, we assume that all atoms are sufficiently close to one another, so that superradiance occurs. The total photon emission $I(t)$ is then given by the sum of photon emission rates for state $\ket{J,M}$, $\Gamma_{M,M-1}$ \eqref{eq:Dicke Decay Rate}, weighted by respective time-dependent probabilities $p_M \left( t \right)$
%%%
\begin{equation}
I \left( t \right) = \sum\limits^{J}_{M=-J+1} p_M \left( t \right) \Gamma_{M,M-1}.
\label{eq:Intensity}
\end{equation}
%%%
A set of rate equations for the probabilities is set up by computing the decay rate for each collective atomic state $\ket{J,M}$ \eqref{eq:Dicke Decay Rate}
%%%
\begin{equation}
\dot{p}_M \left( t \right) = - \Gamma_{M,M-1} p_M \left( t \right) + \Gamma_{M+1,M} p_{M+1} \left( t \right)
\label{eq:Rate Equations}
\end{equation}
%%%
for $-J \leq M \leq J$. Figure \ref{fig:Intensity} shows the emitted intensity for an atomic ensemble of $N = 50$ and $N = 100$ atoms at a distance of $z_{\textrm{A}}=10^{-7} \; \textrm{m}$ from the surface scaled by the single-atom vacuum decay rate $\Gamma^{(0)}$. The superradiant emission burst is very pronounced especially for a large number of atoms. The presence of the surface enhances this effect even further thus giving a Purcell--Dicke enhancement. The insets shows a logarithmic plot of the peak height and the peak width as function of the number of atoms $N$. We observe the well-known properties for the peak height of $\propto N^2$ and the peak width of $\propto 1/N$.

The collective emission by the atomic ensemble is intrinsically related to the Casimir--Polder potential of the atomic ensemble, where we further allow for the presence of a monochromatic coherent driving laser field $\vec{E} \left( \vec{r}_{\textrm{A}}, t \right) = \vec{E} \left( \vec{r}_{\textrm{A}} \right) \cos \left( \omega_{\textrm{L}} t \right)$ of frequency $\omega_{\textrm{L}}$. Based on the approach using the interaction Hamiltonian \eqref{eq:Interaction Hamiltonian} in the framework of macroscopic QED, the time-dependent total Casimir--Polder potential for the atomic cloud reads \cite{Buhmann_Book_2}
%%%
\begin{multline}
U \left( \vec{r}_{\textrm{A}}, t \right) = - \frac{\mi \mu_0}{2 \pi} \int\limits^{\infty}_0 \dif \omega \omega^2 \int\limits^t_0 \dif \tau \me^{-\mi \omega \left( t-\tau \right)}\\
\times \langle \hat{\vec{d}} \left( t \right) \sprod \textrm{Im} \tens{G} \left( \vec{r}_{\textrm{A}}, \vec{r}_{\textrm{A}}, \omega \right) \sprod \hat{\vec{d}}\left( \tau \right) \rangle + \textrm{h.c.}
\label{eq:Casimir Polder Potential}
\end{multline}
%%%
with the dipole moment operator in the Dicke picture $\hat{\vec{d}} \left( t \right) = \sum_{K=M\pm1} \vec{d}_{M,K} \hat{A}_{M,K} \left( t \right)$ being given in terms of collective atomic flip operators $\hat{A}_{M,M-1} = \ket{M} \bra{M-1}$. The differential equation of the expectation value of the nondiagonal elements of the atomic flip operator contains the driving laser field and reads
%%%
\begin{multline}
\langle \dot{\hat{A}}_{M,N} \left( t \right) \rangle = \mi \left[ M-N \right] \omega_{\textrm{A}} \langle \hat{A}_{M,N} \left( t \right) \rangle\\
+ \frac{\mi}{\hbar} \sum\limits_{K=N \pm 1} \vec{E} \left( \vec{r}_{\textrm{A}}, t \right) \sprod \vec{d}_{N,K} \langle \hat{A}_{M,K} \left( t \right) \rangle\\
- \frac{\mi}{\hbar} \sum\limits_{K=M \pm 1} \vec{E} \left( \vec{r}_{\textrm{A}}, t \right) \sprod \vec{d}_{K, M} \langle \hat{A}_{K,N} \left( t \right) \rangle\\
+ \left[ \Gamma_{M+1,M,N,N+1} + \Gamma_{N,N+1,M+1,M} \right] \langle \hat{A}_{M+1,N+1} \left( t \right) \rangle\\
- \left[ \Gamma_{N,N-1,N-1,N} + \Gamma_{M-1,M,M,M-1} \right] \langle \hat{A}_{M,N} \left( t \right) \rangle.
\label{eq:Atomic Flip Operator Nondiagonal}
\end{multline}
%%%
The respective differential equation of the diagonal elements of the atomic flip operator $\langle \hat{A}_{M,M} \left( t \right) \rangle$ is identical with the rate equation for the probability $p_M \left( t \right)$ \eqref{eq:Rate Equations} with the additional term from the driving laser field in Eq.~\eqref{eq:Atomic Flip Operator Nondiagonal}. The decay rates having four indices are defined by $\Gamma_{M,K,N,L} =  \frac{2 \mu_0}{\hbar} \omega^2_{\textrm{A}} \vec{d}_{M,K} \sprod \textrm{Im} \tens{G} \left( \vec{r}_{\textrm{A}}, \vec{r}_{\textrm{A}}, \omega_{\textrm{A}} \right) \sprod \vec{d}_{N,L}$. To calculate the laser-driven Casimir--Polder potential the expectation value of correlated atomic dipole moments $\langle \hat{\vec{d}} \left( t \right) \hat{\vec{d}} \left( \tau \right) \rangle$ is required. In the absence of a driving electric field the potential term \eqref{eq:Casimir Polder Potential} is readily calculated using the quantum regression theorem \cite{Buhmann_Book_2} and the residue theorem. Eventually, the laser-driven Casimir--Polder potential has nonresonant parts, which are not considered in our analysis. The remaining resonant potential represents a sum of potentials for each energy level weighted by the respective probabilities
%%%
\begin{equation}
U \left( \vec{r}_{\textrm{A}}, t \right) = \sum\limits^J_{M=-J+1} p_M \left( t \right) U_M \left( \vec{r}_{\textrm{A}} \right)
\label{eq:Casimir Polder Potential Result}
\end{equation}
%%%
with
%%%
\begin{align}
U_M \left( \vec{r}_{\textrm{A}}, t \right) &= \left( J+M \right) \left( J-M+1 \right) U \left( \vec{r}_i \right),\\
U \left( \vec{r}_i \right) &= -\mu_0 \omega^2_{\textrm{L}} \vec{d} \sprod \textrm{Re} \tens{G}^{(1)} \left( \vec{r}_i, \vec{r}_i, \omega_{\textrm{A}} \right) \sprod \vec{d}^*.
\end{align}
%%%
All damping terms are assumed to be much smaller than the laser frequency and the transition frequency: $\Gamma \ll \omega_{\textrm{L}}, \omega$. The probabilities can also be computed using a master equation \cite{Sinha:2018} with a system Hamiltonian consisting of the atomic Hamiltonian $\hat{H}_{\textrm{A}}$ and the driving Hamiltonian, which is explained in Ref.~\cite{Breuer_Book}. The Lindblad Liouvillian describes spontaneous emission with the collective operators \eqref{eq:Collective Operators}.

\begin{figure}[t!]
	\centering
		\includegraphics[width=\columnwidth]{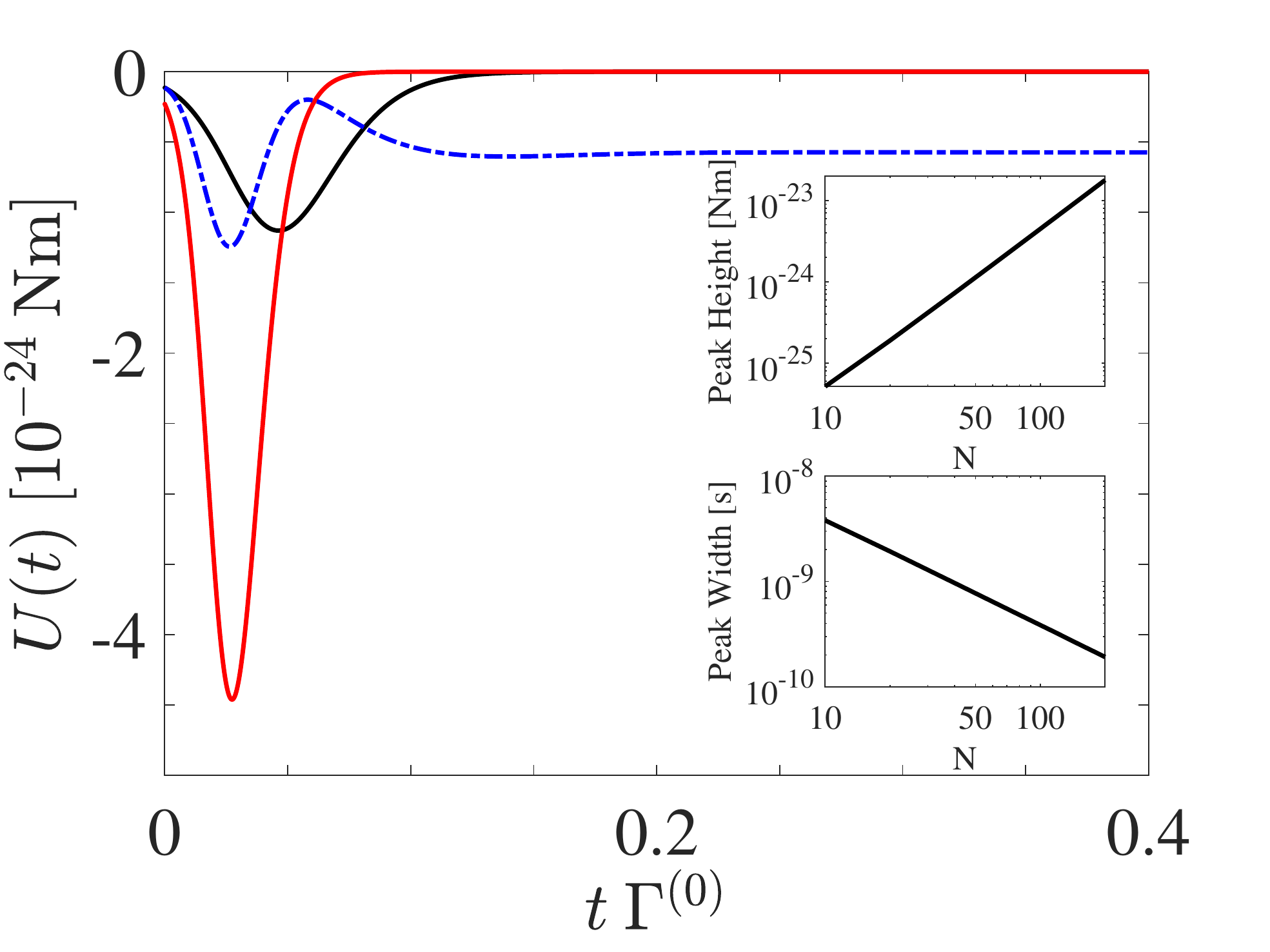}
	\caption{Casimir--Polder potential of an atomic ensemble of $N = 50$ atoms (black line) and $N = 100$ atoms (red line) for Rb atoms with atomic transition frequency of $\omega_{\textrm{A}} = 2.37 \times 10^{15} \; \textrm{rad}/\textrm{s}$ and dipole moment $\left| \vec{d} \right| = 2.53 \times 10^{-29} \; \textrm{Cm}$.  All atoms are located at $x_{\textrm{A}} = 0, \; z_{\textrm{A}} = 10^{-7} \: \textrm{m}$ and are polarized along the $z$-direction. The blue dot-dashed line shows the Casimir--Polder peak for a laser-driven ensemble of $N = 50$ atoms with an intensity of $I = 30000 \: \textrm{W}/\textrm{m}^2$ and a detuning between the laser frequency $\omega_{\textrm{L}}$ and the atomic frequency $\omega$ of $\Delta = \omega_{\textrm{L}} - \omega_{\textrm{A}} = 2 \pi \times 10^8 \; \textrm{rad}/\textrm{s}$, which gives physical results only in the regime $\Gamma t \ll 1$. The insets show the peak height and the full width at half maximum (FWHM) as a function of the number of atoms $N$. We obtain the relations: peak height $\propto N^{1.958}$ and FWHM $\propto N^{-0.999}$.}
	\label{fig:Potential}
\end{figure}

Figure \ref{fig:Potential} shows the collective Casimir--Polder potential \eqref{eq:Casimir Polder Potential Result} of an atomic ensemble of $N=50$ and $N=100$ atoms. As shown in the insets, the peak height and the peak width exhibit the typical superradiant scalings with $N^2$ and $1/N$, respectively and the peak position is thus given by $t=1/\left( N \Gamma \right)$.

By applying an electric driving field, the time scale is additionally governed by the Rabi frequency $\Omega = \vec{d} \sprod \vec{E} \left( \vec{r}_{\textrm{A}} \right)/\hbar$. If $N \Gamma > \Omega$, a pronounced peak remains, the Rabi oscillations are not visible and the curve resembles that without applied electric field. In case of $N \Gamma < \Omega$, Rabi oscillations are superposed on the peak. Figure \ref{fig:Potential} shows the Casimir--Polder potential with applied electric field in the regime $N \Gamma \approx \Omega$, where the peak structure is significantly altered.

In this Letter, we have studied the enhancement of the photon emission and the Casimir--Polder potential due to collective effects and the presence of a surface in a combined Purcell--Dicke effect. We show the connection of the decay rate in the Dicke picture with the single-atom rate via the atomic dipole moments using the symmetric Dicke states. The long-wavelength approximation according to the Dicke model is checked by comparing the joint decay rate of two atoms placed slightly away from each other with the respective Dicke decay rate. The enhancement due to the presence of the surface is described in the form of a Purcell fidelity given by the surface-induced decay rate $\Gamma^{(1)}$ relative to the free-space decay rate $\Gamma^{(0)}$. This enhancement effect is depicted for the total photon emission and the collective Casimir--Polder potential for a mesoscopic number of atoms in the vicinity of a perfectly conducting mirror, showing peak heights and widths which fulfill the criteria of superradiance. As shown, an external driving laser can be used to manipulate the dynamics of the potential.

The Dicke enhancement of the Casimir--Polder force can be exploited to significantly increase sensitivity without having to extend interaction times. In this way, such forces can be used as sensitive probes of atomic or surface properties such as chirality \cite{Barcellona:2017} or CP violation \cite{Buhmann:2018} or even facilitate the (spectroscopic) detection of quantum friction \cite{Milton:2016, Klatt:2016}.

We acknowledge helpful discussions with Miguel Bastarrachea, Robert Bennett, Diego Dalvit, Francesco Intravaia, Bj\"{o}rn Kubala and Ian Walmsley. This work was supported by the German Research Foundation (DFG, Grants BU 1803/3-1 and GRK 2079/1). S.Y.B is grateful for support by the Freiburg Institute of Advanced Studies.

%%%%

%merlin.mbs apsrev4-1.bst 2010-07-25 4.21a (PWD, AO, DPC) hacked
%Control: key (0)
%Control: author (72) initials jnrlst
%Control: editor formatted (1) identically to author
%Control: production of article title (-1) disabled
%Control: page (0) single
%Control: year (1) truncated
%Control: production of eprint (0) enabled
%

\end{document}